\documentclass[runningheads]{miccai_llncs}

\usepackage{graphicx}
\usepackage[toc]{glossaries}
\usepackage[hidelinks]{hyperref}
\hypersetup{
    colorlinks=true,
    linkcolor=blue,
    filecolor=magenta,      
    urlcolor=cyan,
}

\usepackage{amsmath}
\usepackage{amssymb}
\usepackage{bbm}
\usepackage{eucal}
\usepackage{enumitem}
\usepackage{booktabs}
\usepackage{caption,subcaption}
\usepackage{bbm}
\usepackage{tikz}
\usepackage{lipsum}
\usepackage{adjustbox}

\DeclareMathAlphabet\mathbfcal{OMS}{cmsy}{b}{n}

\usepackage{tcolorbox}
\usepackage{nicefrac,xfrac}

\DeclareMathOperator*{\relu}{\textsc{ReLU}}

\newcommand{\figref}[1]{Figure~\ref{#1}}
\newcommand{\tableref}[1]{Table~\ref{#1}}
\renewcommand{\eqref}[1]{Equation~\ref{#1}}

\newcommand{\mycomment}[1]{}

\usepackage{array}
\usepackage{multirow}
\newcolumntype{L}[1]{>{\raggedright\let\newline\\\arraybackslash\hspace{0pt}}m{#1}}
\newcolumntype{C}[1]{>{\centering\let\newline\\\arraybackslash\hspace{0pt}}m{#1}}
\newcolumntype{R}[1]{>{\raggedleft\let\newline\\\arraybackslash\hspace{0pt}}m{#1}}

\makeglossaries

\newacronym[plural=CNNs,firstplural=Convolutional Neural Networks (CNNs)]{cnn}{CNN}{Convolutional Neural Network}
\newacronym{dl}{DL}{deep learning}
\newacronym{qcmlb}{QC-MLB}{Question-Centric Multimodal Low-rank Bilinear}
\newacronym{bert}{BERT}{Bidirectional Encoder Representations from Transformers}
\newacronym{bleu}{BLEU}{Bilingual Evaluation Understudy}
\newacronym{mlm}{MLM}{Masked Language Model}
\newacronym{nsp}{NSP}{Next Sentence Prediction}
\newacronym{relu}{ReLU}{rectified linear unit}
\newacronym{nn}{NN}{neural network}
\newacronym{chal}{ImageCLEF-VQA-Med}{ImageCLEF-VQA-Med}
\newacronym{proposed}{\textit{\textless~Model~\textgreater}}{\textbf{Full name of the proposed model}}
\newacronym{mri}{MRI}{magnetic resonance imaging}
\newacronym{brats}{BraTS19}{Brain Tumors in Multimodal Magnetic Resonance Imaging Challenge 2019}
\newacronym{kits}{KiTS19}{Kidney Tumor Segmentation Challenge 2019}
\newacronym{ibsr}{IBSR18}{Internet Brain Segmentation Repository}
\newacronym{hene}{HeNe}{Ume{\aa} Head and Neck}
\newacronym{pros}{ProST}{Ume{\aa} Prostate}

\newacronym{ct}{CT}{computed tomography}
\newacronym{t1c}{T1c}{post-contrast T1-weighted}
\newacronym{t2}{T2w}{T2-weighted}
\newacronym{t1}{T1w}{T1-weighted}
\newacronym{flair}{FLAIR}{T2  Fluid  Attenuated  Inversion  Recovery}
\newacronym{lgg}{LGG}{low grade glioma}
\newacronym{hgg}{HGG}{high grade glioma}

\newacronym{svm}{SVMs}{Support-vector Machines}
\newacronym{crf}{CRF}{Conditional Random Field}
\newacronym{vae}{VAE}{Variational Auto-Encoder}
\newacronym{tunet}{TuNet}{End-to-end Hierarchical Tumor Segmentation using Cascaded Networks}
\newacronym{dram}{DRAM}{dynamic random access memory}
\newacronym{gpus}{GPUs}{graphics processing units}

\newacronym{hpc2n}{HPC2N}{High Performance Computer Center North}

\newacronym{dsc}{DSC}{S{\o}rensen-Dice coefficient}
\newacronym{se}{SE}{standard error}
\newacronym{seb}{SEB}{Squeeze-and-Excitation block}
\newacronym[plural=REBs,firstplural=ResNet blocks]{res}{REB}{ResNet block}
\newacronym{hd95}{HD95}{$95^{th}$ percentile of the Hausdorff distance}
\newacronym{dauc}{DAUC}{Dice Area Under Curve}
\newacronym{rftp}{RFTPs}{Ratio of Filtered True Positives}

\newacronym[plural=SDs,firstplural=standard deviations (SDs)]{sd}{SD}{standard deviation}
\usepackage{scalefnt}

\begin{document}
\title{TuNet: End-to-end Hierarchical Brain Tumor Segmentation using Cascaded Networks}

\author{Minh H. Vu\inst{1} \and Tufve Nyholm\inst{1} \and Tommy L\"{o}fstedt\inst{1}}

\institute{Department of Radiation Sciences, Ume{\aa} University, Ume{\aa}, Sweden \\ \email{minh.vu@umu.se}}
\maketitle

\begin{abstract}


Glioma is one of the most common types of brain tumors; it arises in the glial cells in the human brain and in the spinal cord. In addition to having a high mortality rate, glioma treatment is also very expensive. Hence, automatic and accurate segmentation and measurement from the early stages are critical in order to prolong the survival rates of the patients and to reduce the costs of the treatment. In the present work, we propose a novel end-to-end cascaded network for semantic segmentation that utilizes the hierarchical structure of the tumor sub-regions with ResNet-like blocks and Squeeze-and-Excitation modules after each convolution and concatenation block. By utilizing cross-validation, an average ensemble technique, and a simple post-processing technique, we obtained dice scores of 88.06, 80.84, and 80.29, and Hausdorff Distances (95th percentile) of 6.10, 5.17, and 2.21 for the whole tumor, tumor core, and enhancing tumor, respectively, on the online test set. 


\end{abstract}

\section{Introduction}
\label{sec:intro}

Glioma is among the most aggressive and dangerous types of cancer~\cite{holland2001progenitor}, leading, for instance, to around 80~\% and 75~\% of all malignant brain tumors diagnosed in the United States~\cite{gholami2016inverse} and Sweden~\cite{asklund2012overlevnanden}, respectively. Gliomas with different prognosis and numerous heterogeneous histological sub-regions, such as edema, necrotic core, and enhancing and non-enhancing tumor core, are classified into four world health organisation (WHO) grades according to their aggressiveness: \gls{lgg} (class I and II, considered as slow-growing), and \gls{hgg} (class III and IV, considered as fast-growing). 


The aim of the \gls{brats}~\cite{menze2014multimodal,bakas2017advancing,bakas2018identifying,bakas2017segmentation,bakas2017segmentation17} is in evaluating and finding new state-of-the-art methods for tumor segmentation and to set a gold standard for the segmentation of intrinsically heterogeneous brain tumors. \gls{brats} provided a large data set comprising multi-institutional pre-operative \gls{mri} scans with four sequences: \gls{t1c}, \gls{t2}, \gls{t1}, and \gls{flair}. Masks were annotated manually by one to four raters followed by improvements by expert raters. The segmentation performance of the participants was measured using \gls{dsc}, sensitivity, specificity, and \gls{hd95}.

\begin{figure}[!t]
    \begin{center}
        \includegraphics[width=0.8\textwidth]{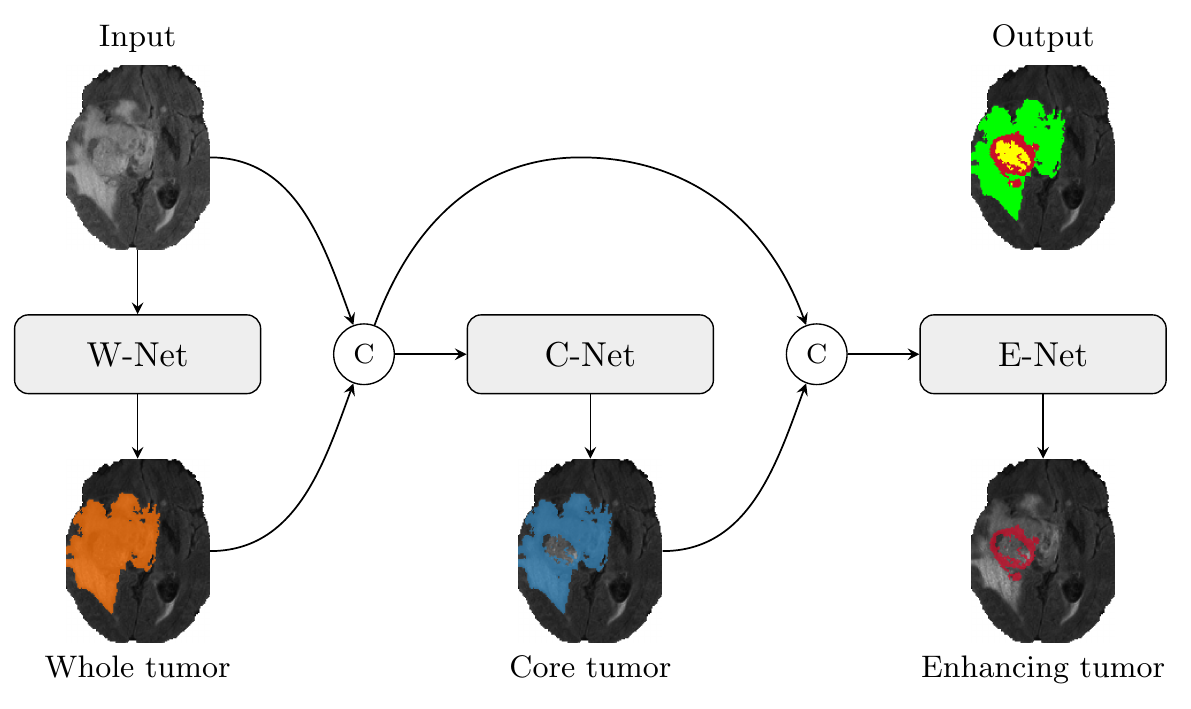}
    \end{center}
    \caption[]{Schematic visualization of TuNet. The input is the \gls{mri} volumes as four different channels, while the output is the predicted masks for the three labels: the necrotic and non-enhancing tumor core (NCR/NET---label 1, green), the peritumoral edema (ED---label 2, yellow) and the GD-enhancing tumor (ET---label 3, red). First, the multi-modal input is fed into the W-Net to generate a probability map of the whole tumor region (orange) including labels 1, 2 and 3. Second, the concatenation of the whole tumor probability map and the input is passed through the C-Net to produce the tumor core probability map (blue) including labels 1 and 3.
    Third, the two obtained maps from W-Net and C-Net are concatenated with the multi-modal input and then fed into the E-Net to generate an enhancing tumor probability map including label 3. Last, the outputs of W-Net, C-Net, and E-Net are merged to produce the final brain tumor mask.
    }
    \label{fig:method}
\end{figure}

Traditional discriminative approaches, such as 
Support-vector Machines~\cite{boser1992training}, have been widely used in medical image segmentation. In recent years, \glspl{cnn} have achieved state-of-the-art performance in numerous computer vision tasks.
In the field of medical image segmentation, a fully convolutional encoder-decoder neural network named U-Net, introduced by Ronneberger \textit{et al.}~\cite{ronneberger2015unet}, has received a lot of attention in recent years. With its success, it is unsurprising that the U-Net has motivated many top-ranking teams in segmentation competitions in previous years~\cite{isensee2018no,chen2018encoder,Myronenko2018,Wang_2018}.


\begin{figure}[!t]
    \begin{center}
        \includegraphics[width=0.9\textwidth]{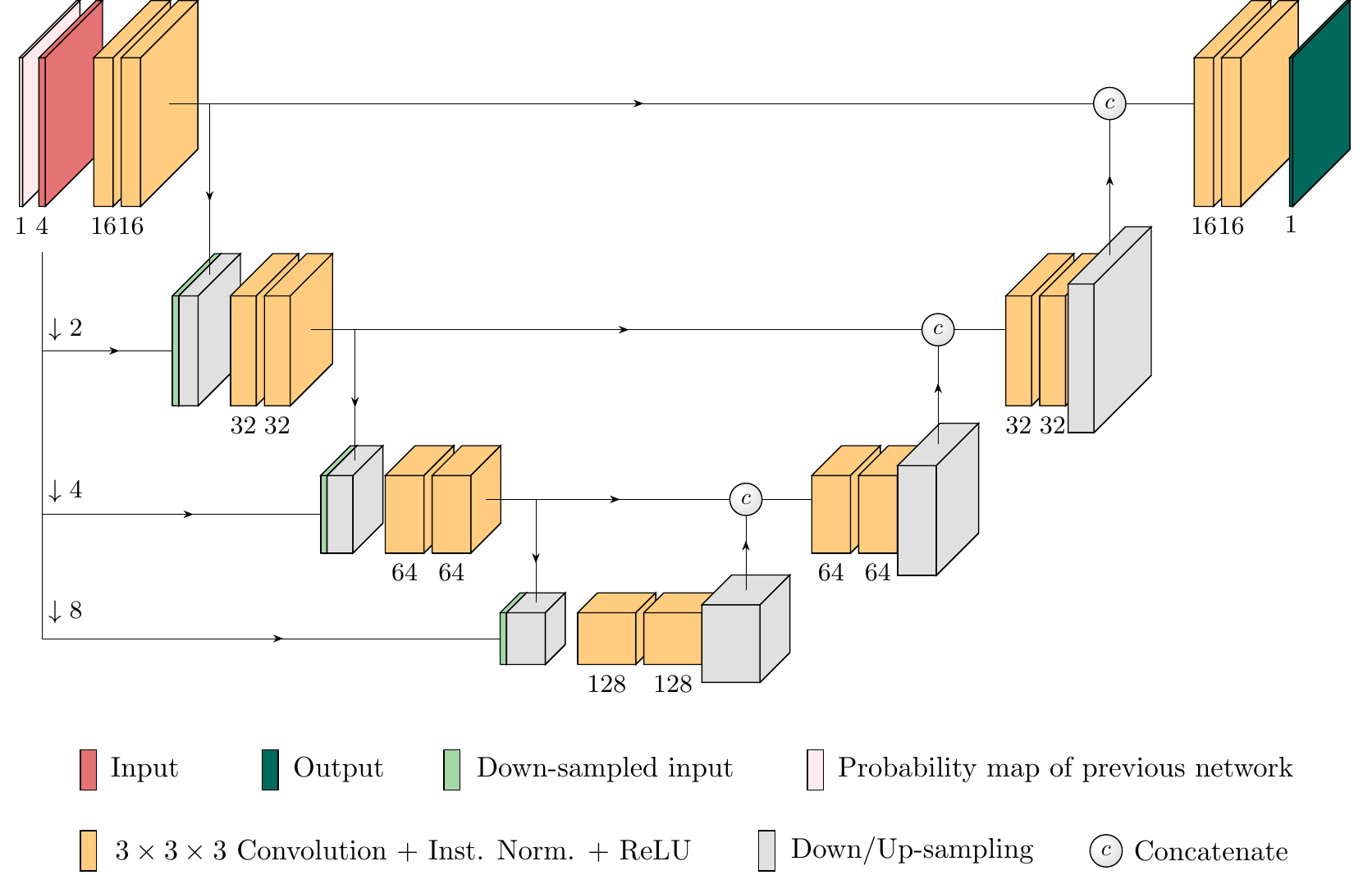}
    \end{center}
    \caption[]{Our C-Net and E-Net with a U-Net backbone. The W-Net does not include the probability map produced by a previous sub-network. To enrich the feature map at each level in the encoder part, we concatenate the down-sampled original input (light green) with the down-sampled output (grey) of the convolution block (yellow). Each convolution block comprises a convolution operation, an instance normalization followed by a $\relu$ activation function. 
    The W-Net constrains the C-Net, while the C-Net constrains the E-Net. 
    The E-Net functions as a regularizer for the C-net, while the C-Net plays the same role for the W-Net.
    }
    \label{fig:unet}
\end{figure}

Kitrungrotsakul \textit{et al.}~\cite{kitrungrotsakul2019cascade} presented CasDetNet-CLSTM to detect mitotic events in 4D microscopic images by introducing a connection between a region-based convolutional neural network and convolutional long short-term memory. In the BraTS18, instead of proposing a new network, Isensee \textit{et al.}~\cite{isensee2018no} focused on the training process, and made only minor modifications to the U-Net, used additional training data, and applied a simple post-processing technique. Myronenko~\cite{Myronenko2018} was the winner of BraTS18 by introducing a U-Net-like network with an asymmetrically large encoder and a \gls{vae} branch to reconstruct the input image to add guidance and regularization to the encoder part of the network.

In another work, Wang \textit{et al.}~\cite{Wang_2018} proposed an encoder-decoder cascaded anisotropic \gls{cnn}, that won the second place in the BraTS17, that hierarchically segments the whole tumor, tumor core, and enhancing tumor core sequentially by separating the complex brain tumor segmentation problem into three binary segmentation problems. We argue that: (1) the training process employed in~\cite{Wang_2018} might be time-consuming as there are three separate binary segmentation problems, and (2) the lack of regularization could lead to overfitting.



Motivated by the successes of the cascaded anisotropic network, introduced in~\cite{Wang_2018}, and the \gls{vae} branch, presented in~\cite{Myronenko2018}, we propose a novel architecture, denoted \gls{tunet}. \gls{tunet} exploits separating a complex problem into less challenging sub-problems, and also attempts to avoid overfitting by using heavy regularization through multi-task learning. In the present work, we constructed three U-Net-like networks, each was used to segment a specific tumor region, \textit{i.e.} whole tumor, core tumor, and enhancing tumor. We connected the three sub-networks to form an end-to-end cascaded network in the hope that three sub-networks could mutually regularize each other to prevent overfitting and reduce the training time.

\section{Methods}
\label{sec:method}

Motivated by the drawbacks in~\cite{Wang_2018} and the \gls{vae} branch in~\cite{Myronenko2018}, we thus propose a framework that we expect not only utilizes region-based optimization, but we also hypothesise that it will prevent or reduce overfitting.


\subsection{Cascaded Architecture}
\label{subsec:framework}

\figref{fig:method} illustrates the proposed end-to-end cascaded architecture. As in~\cite{Wang_2018}, we employed three encoder-decoder networks to cope with the three aforementioned tumor regions. We denote them W-Net (whole tumor network), C-Net (core tumor network), and E-Net (enhancing tumor network). In our proposed approach, instead of using three separate networks, we joined them together to form an end-to-end cascaded network. 




\subsection{Segmentation Network}
\label{subsec:cnn}



\figref{fig:unet} shows the proposed symmetric encoder-decoder C-Net and E-Net. The patch size was set to $80 \times 96 \times 64$. Note that the probability map (pink) was concatenated with the input image as a first step for the C-Net and the E-Net, but not for the W-Net. We also concatenated the max-pooled feature maps with the down-sampled input (light green) in order to enhance the feature maps at the beginning of each level in the encoder part. The base number of filters was set to 16 and the number of filters were doubled at each level. Skip-connections were used like in the U-Net. The probability maps produced at the end of the decoder part of each sub-network had the same spatial size as the original image and were activated by a logistic sigmoid function.

\subsection{ResNet Block}
\label{subsec:res}

He \textit{et al.}~\cite{he2016deep} proposed the ResNet as a way to make training deep networks easier, and won the 1$^\text{st}$ place in the ImageNet Large Scale Visual Recognition Challenge (ILSVRC) 2015 classification task. It has been shown that ResNet-like networks are not only easier to optimize but also boost accuracy from considerably deeper networks~\cite{he2016deep,Myronenko2018}. Inspired by the winner of BraTS18 \cite{Myronenko2018}, we replaced the convolution blocks used in the traditional U-Net by \glspl{res} in two of our models (see \figref{fig:unet} and \tableref{tab:cv}). An illustration of the ResNet-like block can be seen in \figref{fig:compare}.

\subsection{Squeeze-and-Excitation Block}
\label{subsec:se}

\begin{figure}[!t]
    \begin{center}
        \includegraphics[width=0.9\textwidth]{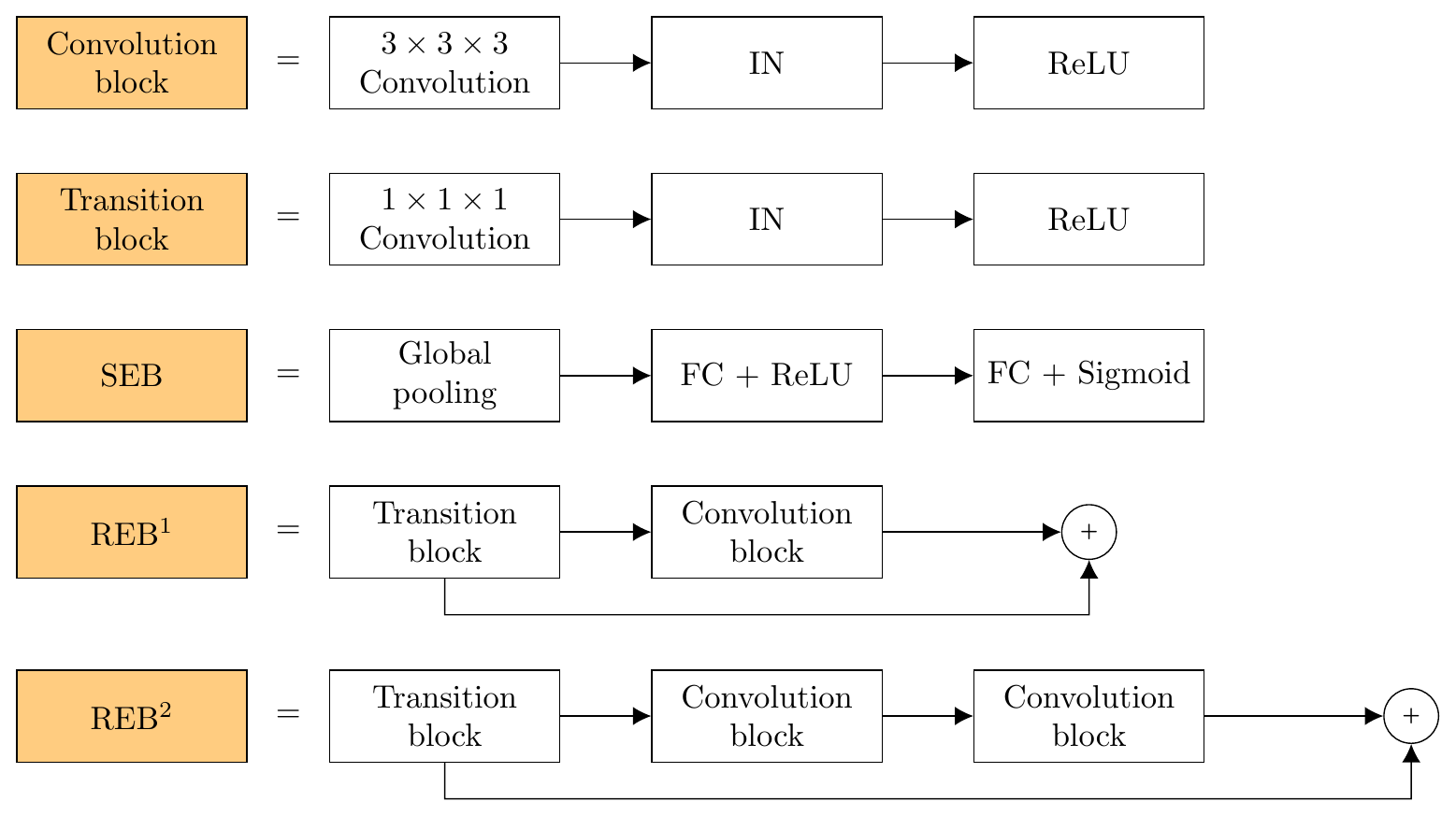}
    \end{center}
    \caption[]{The convolution blocks used in the different experiments (see \tableref{tab:cv}). \gls{seb}, \gls{res}$^1$ and \gls{res}$^2$ denote \glsdesc{seb} and two variations of \acrlongpl{res}, respectively. Here, IN and FC stand for instance normalization and fully-connected layer, respectively; while rectified linear unit (ReLU) and Sigmoid are activation functions.}
    \label{fig:compare}
\end{figure}

We added a \gls{seb} as developed by Hu \textit{et al.}~\cite{hu2018squeeze}
after each convolution and concatenation block. The \gls{seb} is a computationally efficient means to incorporate channel-wise inter-dependencies, and has been widely used to improve network performances by significant margins~\cite{hu2018squeeze}. \gls{seb} is also illustrated in \figref{fig:compare}.


\subsection{Preprocessing and Augmentation}
\label{subsec:prep}

We normalized all input images to have mean zero and unit variance. To increase the data set size, we employed simple on-the-fly data augmentation by randomly rotating the images within a range of $-1$ to $1$ degrees and random mirror flips (on the $x$-axis) with a probability of 0.5. We also experimented with median denoising, but it did not demonstrate any additional improvements.


\subsection{Post-processing}
\label{subsec:post}

One of the most difficult tasks of \gls{brats} is to detect small vessels in the tumor core and to label them as edema or as necrosis. To cope with the fact that \gls{lgg} patients may have no enhancing tumor region, Isensee \textit{et al.}~\cite{isensee2018no} proposed to replace all enhancing tumor regions with less than 500 voxels by necrosis. We employed that approach
in the present work as well.
Additionally, we kept decreasing the threshold segmentation from 0.5 to 0.3, 0.1, 0.05, and 0.01, respectively, if no core tumor region was found. This technique resolved several cases where the networks failed to distinguish between core and whole tumor.

\subsection{Ensemble of Multiple Models}
\label{subsec:ensemble}

We employed five-fold cross-validation when training on the 335 cases (259 \gls{hgg} + 76 \gls{lgg}) of \gls{brats}, and did not use any additional in-house data.
We trained 30 models in total and used 15 of them, that correspond to \gls{tunet} + \gls{seb}, \gls{tunet} + \gls{res}$^1$ + \gls{seb} and \gls{tunet} + \gls{res}$^2$ + \gls{seb}, to cast votes when predicting the final label maps. We used an average ensemble approach, such that
\begin{equation} \label{eqn:ensemble_avg}
    p_{c} = \frac{1}{M} \sum_{m=1}^{M} f_{mc},
\end{equation}
where $p_c \in \mathbb{R}^{|C|}$ and $f_{mc} \in \mathbb{R}^{|C|}$
denote the final probability of label $c$ and the probability of label $c$ generated by model $m=1,\ldots,M$ at an arbitrary voxel, respectively. Here, $C=\{1, 2, 3\}$ is the set of tumor labels, and thus $c \in C$.

The ensemble that we used was computed as the average of the prediction outputs of \gls{tunet} + \gls{seb}, \gls{tunet} + \gls{res}$^1$ + \gls{seb} and \gls{tunet} + \gls{res}$^2$ + \gls{seb} (see \tableref{tab:cv} and \tableref{tab:valid}).


\subsection{Task 3: Quantification of Uncertainty in Segmentation}
\label{subsec:task3}


In \gls{brats}, the organizers decided to include a new task that focuses on exploring uncertainty measures in the context of tumor segmentation on three glioma regions: whole, core, and enhancing. This task, called ``Quantification of Uncertainty in Segmentation'', aimed at rewarding participating methods with resulting predictions that are: (a) confident when correct and (b) uncertain when incorrect. Participants were called on to generate uncertainty maps associated with the resulting labels at every voxel with values in $[0, 100]$, where 0 represents the most certain prediction and 100 represents the most uncertain.

Our proposed approach, \gls{tunet}, benefits from this task since, fundamentally, it segments the brain tumor \gls{mri} images into three aforementioned tumor regions instead of partitioning into three labels. We define an uncertainty score, $u^r_{i,j,k}$, at voxel $(i,j,k)$ as:
\begin{equation}
    u^r_{i,j,k} =   
    \begin{cases} 
        200 (1-p^{r}_{i,j,k}) & \text{if } p^{r}_{i,j,k} \geq 0.5 \\
        200 p^{r}_{i,j,k}     & \text{if } p^{r}_{i,j,k} < 0.5
  \end{cases}
\end{equation}
where thus $u^r_{i,j,k} \in [0, 100]^{|\mathcal{R}|}$ and $p^{r}_{i,j,k} \in [0, 1]^{|\mathcal{R}|}$
denote the uncertainty score map and probability map (the network's likelihood outputs) corresponding to tumor region, $r \in \mathcal{R}$, as produced by the TuNet (see \figref{fig:method}), where $\mathcal{R}$ is the set of tumor regions, \textit{i.e.} whole, core, and enhancing region.


\section{Experiments}
\label{subsec:exp}

We implemented our network in  Keras 2.2.4\footnote{\url{https://keras.io}} using TensorFlow 1.12.0\footnote{\url{https://tensorflow.org}} as the backend. This research was conducted using the resources of the High Performance Computing Center North (HPC2N)\footnote{\url{https://www.hpc2n.umu.se/}} at Ume{\aa} University, Ume{\aa}, Sweden, and the experiments were run on NVIDIA Tesla V100 16~GB GPUs.

We report the results on the \gls{brats} training set (335 cases) using cross-validation and on the validation set (125 cases) by uploading our predicted masks and corresponding uncertainty maps to the evaluation server. The evaluation metrics of Task 1---Segmentation included: \gls{dsc}, sensitivity, specificity, and \gls{hd95}; while the evaluation metrics of Task 3---Quantification of Uncertainty in Segmentation were \gls{dauc} and \gls{rftp}.


\subsection{Implementation Details and Training}
\label{subsec:train}


For evaluation of the segmentation performance, we employed the \gls{dsc}, defined as
\begin{equation}
    D(X,Y)=\frac{2 |X \cap Y|}{|X| + |Y|},
\end{equation}
where $X$  and $Y$ denote the output segmentation and its corresponding ground truth, respectively.

The \gls{hd95} is defined as the largest value in the set of closest distances between two structures, or
\begin{equation}
    H(X,Y)=\max\big\{ \max_{x\in X} \min_{y \in Y} d(x,y), \max_{y\in Y} \min_{x \in X} d(y,x) \big\},
\end{equation}
where $d(x,y)$ denotes the Euclidian distance between two points $x \in X$ and $y \in Y$. It is common practice to report the $95^{th}$ percentile instead of the maximum to compensate for outliers.

The loss function used for training contained three terms,
\begin{equation}
    \mathcal{L}(u, v) = \mathcal{L}_{\textit{whole}}(u, v) + \mathcal{L}_{\textit{core}}(u, v) + \mathcal{L}_{\textit{enh}}(u, v),
\end{equation}
where $\mathcal{L}_{\textit{whole}}$,  $\mathcal{L}_{\textit{core}}$, and $\mathcal{L}_{\textit{enh}}$ where the soft dice loss of whole, core, and enhancing tumor regions, respectively, and where the soft dice loss is defined as
\begin{equation}
    \mathcal{L}_{dice}(u, v) = \frac{-2 \sum_i u_i v_i}{\sum_i u_i + \sum_i v_i + \epsilon},
\end{equation}
in which $u$ is the softmax output of the network, $v$ is a one-hot encoding
of the ground truth segmentation map, and $\epsilon$ is a small constant added to avoid division by zero.

We used the Adam optimizer~\cite{kingma2014adam} with a learning rate of $1 \cdot 10^{-4}$ and momentum parameters of $\beta_1=0.9$ and $\beta_2=0.999$. We also used $L_2$ regularization with a penalty parameter of $1 \cdot 10^{-5}$, that was applied to the kernel weight matrices, for all convolutional layers to cope with overfitting. The activation function of the final layer was the logistic sigmoid function.

All models were trained for 200 epochs, with a mini-batch size of four. To prevent over-fitting, we selected a patience period and dropped the learning rate by a factor of $0.2$ if the validation loss did not improve over six epochs. Further, the training process was stopped if the validation loss did not improve after $15$ epochs. The training time for a single model was around 35 hours on an NVIDIA Tesla V100 GPU.

\begin{table}[!t]
    \def\width{1 cm}
    \def\widthdetail{3.7 cm}
    \caption{Mean \gls{dsc} (higher is better) and \gls{hd95} (lower is better) and their \glspl{sd} (in parentheses) computed from the five-folds of cross-validation on the training set (335 cases) for the different models. Details of the variations of the convolution blocks used for each model is shown in \figref{fig:compare}.
    }
    \centering
    \begin{tabular}{L{\widthdetail} l C{\width} C{\width} C{\width} l C{\width} C{\width} C{\width}}
    \toprule
                                                && \multicolumn{3}{c}{\gls{dsc}} && \multicolumn{3}{c}{\gls{hd95}} \\
    Model                                       && whole  & core   & enh.   && whole  & core   & enh.   \\
    \cmidrule{1-1}\cmidrule{3-5}\cmidrule{7-9}
    \gls{tunet}                                 && 89.89 (2.07)   & 84.08 (4.02)     & 74.92 (7.24)     && 6.22 (2.66)  & 6.82 (2.93)  & 4.72 (2.21)    \\
    \gls{tunet} + \gls{seb}                     && 91.38 (2.02)   & 85.93 (3.77)     & 78.11 (7.29)     && 4.50 (2.34)  & 6.70 (2.72)  & 4.24 (2.07)    \\
    \gls{tunet} + \gls{res}$^1$                 && 90.68 (1.98)   & 86.07 (3.84)     & 75.88 (7.36)     && 5.14 (2.07)  & \textbf{5.16 (1.99)}  & 4.20 (1.99)    \\
    \gls{tunet} + \gls{res}$^1$ + \gls{seb}     && 90.86 (2.12)   & 86.30 (3.72)     & 76.20 (7.32)     && \textbf{5.10 (2.29)}  & 5.72 (2.62)  & 3.89 (1.61)    \\
    \gls{tunet} + \gls{res}$^2$                 && 90.77 (2.18)   & 85.84 (3.96)     & 75.22 (7.30)     && 5.42 (2.50)  & 6.00 (2.38)  & 4.56 (2.09)    \\
    \gls{tunet} + \gls{res}$^2$ + \gls{seb}     && 91.90 (2.00)   & 86.09 (3.81)     & 77.43 (7.28)     && 5.07 (2.38)  & 6.20 (2.45)  & 3.98 (1.87)    \\    
    \cmidrule{1-1}\cmidrule{3-5}\cmidrule{7-9}
    Ensemble                                    && 91.92 (2.18)   & 86.35 (3.81)     & 78.01 (7.30)     && 5.30 (2.19)  & 5.80 (2.69)  & 3.45 (1.93)     \\
    Ensemble + post-process                     && \textbf{91.92 (2.18)}   & \textbf{86.45 (3.79)}     & \textbf{78.72 (7.10)}     && 5.30 (2.19)  & 5.75 (2.65)  & \textbf{3.08 (1.90)}     \\
    \bottomrule
    \end{tabular}
    \label{tab:cv}
\end{table}

\section{Results and Discussion}
\label{sec:results}

The key strengths of the proposed method are: (1) it takes advantage of the hierarchical structure of the tumor sub-regions, since the whole tumor region must contain the core tumor, and the core tumor region must contain the enhancing tumor region, and (2) it consists of three connected sub-networks that mutually regularize each other, \textit{i.e.} the E-Net functions as a regularizer for the C-net, C-Net plays that role for the W-Net, and the W-Net and the E-Net constrain each other through the C-Net. 

\begin{table}[!t]
\def\width{1. cm}
\def\widthdetail{3.7 cm}
\caption{Results of Segmentation Task on \gls{brats} validation data (125 cases). The results were obtained by computing the mean of predictions of five models trained over the folds. ``UmU'' denotes the name of our team and the ensemble of \gls{tunet} + \gls{seb}, \gls{tunet} + \gls{res}$^1$ + \gls{seb} and \gls{tunet} + \gls{res}$^2$ + \gls{seb}.  The metrics were computed by the online evaluation platform. All the predictions were post-processed before submitting to the server. Bottom rows correspond to the top-ranking teams from the online system.}
\centering
    \begin{tabular}{L{\widthdetail} l C{\width} C{\width} C{\width} l C{\width} C{\width} C{\width}}
    \toprule
                                                && \multicolumn{3}{c}{\gls{dsc}} && \multicolumn{3}{c}{\gls{hd95}} \\
    Model                                       && whole    & core   & enh.     && whole  & core   & enh.   \\
    \cmidrule{1-1}\cmidrule{3-5}\cmidrule{7-9}
    \gls{tunet} + \gls{seb}                     &  & 90.41  & 81.67  & 78.97    && 4.35   & 6.12   & 3.35   \\
    \gls{tunet} + \gls{res}$^1$ + \gls{seb}     &  & 90.06  & 79.43  & 78.12    && 4.66   & 7.98   & 3.34   \\
    \gls{tunet} + \gls{res}$^2$ + \gls{seb}     &  & 90.34  & 79.14  & 77.38    && 4.29   & 8.80   & 3.57   \\    
    \cmidrule{1-1}\cmidrule{3-5}\cmidrule{7-9}
    UmU                                         &  & 90.34 & 81.12 & 78.42      && 4.32      & 6.28  & 3.70 \\    
    \cmidrule{1-1}\cmidrule{3-5}\cmidrule{7-9}
    ANSIR                                       &  & 90.09 & 84.38 & 80.06      && 6.94      & 6.00  & 4.52 \\
    lfn\_                                       &  & 90.91 & 85.48 & 80.24      && 4.35      & \textbf{5.32}  & 3.88 \\
    SVIG1                                       &  & \textbf{91.16} & 85.79 & 81.33      && \textbf{4.10}      & 5.92  & 4.21 \\
    NVDLMED                                     &  & 91.01 & \textbf{86.22} & \textbf{82.28}      && 4.42      & 5.46  & \textbf{3.61} \\
    \bottomrule
\end{tabular}
\label{tab:valid}
\end{table}

\begin{table}[!t]
\def\width{1. cm}
\def\widthdetail{3.7 cm}
\caption{Results of Segmentation Task on \gls{brats} test data (166 cases).}
\centering
    \begin{tabular}{L{\widthdetail} l C{\width} C{\width} C{\width} l C{\width} C{\width} C{\width}}
    \toprule
                                                && \multicolumn{3}{c}{\gls{dsc}} && \multicolumn{3}{c}{\gls{hd95}} \\
    Model                                       && whole    & core   & enh.     && whole  & core   & enh.   \\
    \cmidrule{1-1}\cmidrule{3-5}\cmidrule{7-9}
    UmU                                         &  & 88.06 & 80.84 & 80.29      && 6.10   & 5.17   & 2.20 \\    
    \bottomrule
\end{tabular}
\label{tab:test}
\end{table}

\begin{table}[!t]
\def\width{1. cm}
\def\widthdetail{3.7 cm}
\caption{Results of Quantification of Uncertainty Task on \gls{brats} validation data (125 cases) including mean \gls{dauc} (higher is better) and \gls{rftp} (lower is better). The results were obtained by computing the mean of predictions of five models trained over the folds. ``UmU'' denotes the name of our team and the ensemble of \gls{tunet} + \gls{seb}, \gls{tunet} + \gls{res}$^1$ + \gls{seb} and \gls{tunet} + \gls{res}$^2$ + \gls{seb}.  The metrics were computed by the online evaluation platform. The bottom rows correspond to the top-ranking teams from the online system. The proposed method was ranked among the top that was evaluated on the online test set.}
\centering
    \begin{tabular}{L{\widthdetail} l C{\width} C{\width} C{\width} l C{\width} C{\width} C{\width}}
    \toprule
                                                && \multicolumn{3}{c}{DAUC} && \multicolumn{3}{c}{RFTPs} \\
    Model                                       && whole    & core   & enh.   && whole  & core   & enh.   \\
    \cmidrule{1-1}\cmidrule{3-5}\cmidrule{7-9}
    \gls{tunet} + \gls{seb}                     &  & 87.52  & 79.90  & 75.97  && 4.50   & 12.97  & 6.59   \\
    \gls{tunet} + \gls{res}$^1$ + \gls{seb}     &  & 86.13  & 80.01  & 75.13  && 5.80   & 16.50  & 8.02   \\
    \gls{tunet} + \gls{res}$^2$ + \gls{seb}     &  & 87.40  & 79.32  & 75.90  && 5.50   & 15.30  & 8.53   \\    
    \cmidrule{1-1}\cmidrule{3-5}\cmidrule{7-9}
    UmU                                         &  & 87.47  & 79.86  & 75.89  && 5.83   & 13.72  & 8.46   \\
    \cmidrule{1-1}\cmidrule{3-5}\cmidrule{7-9}
    ANSIR                                       &  & 89.42  & 88.72  & 87.67  && 1.23   & 6.16   & 4.63   \\
    NVDLMED                                     &  & 89.95  & 88.03  & 84.67  && 1.68   & \textbf{2.43}   & 2.86   \\
    TEAM\_ALPACA                                &  & \textbf{90.93}  & \textbf{88.08}  & \textbf{87.90}  && \textbf{1.54}   & 3.82   & \textbf{2.59}   \\
    \bottomrule
\end{tabular}
\label{tab:valid_unc}
\end{table}

\tableref{tab:cv} shows the mean \gls{dsc} and \gls{hd95} scores and \glspl{sd} computed from the five-folds of cross-validation on the training set. As can be seen in \tableref{tab:cv}, with \gls{dsc} of $89.89/84.08/74.92$ (whole/core/enh.) using cross-validation on the training set and $89.70/76.96/77.90$ (whole/core/enh.) on the validation set, our \textit{baseline} (\gls{tunet}) produced acceptable results. Adding \gls{seb} to  \gls{tunet} (\gls{tunet} + \gls{seb}) improved the \gls{dsc} on all tumor regions; however, only core and core+enhancing \gls{dsc} scores were boosted when adding \gls{seb} to two variations, that used ResNet-like blocks, \textit{i.e.} \gls{tunet} + \gls{res}$^1$ + \gls{seb} and \gls{tunet} + \gls{res}$^2$ + \gls{seb}.

We gained a few \gls{dsc} points when the average ensemble technique (Ensemble), \textit{i.e.} \gls{dsc} reached $91.92/86.45/78.72$ (whole/core/enh). The post-processing step only improved slightly the Ensemble model on the core region, but boosted the \gls{dsc} of the enhancing region by a large margin, from 78.01 to 78.72, making it (Ensemble + post-processing) the best-performing model of all proposed models on the training set.

\begin{figure}[!t]
    \def\subfigsize{0.175\textwidth}
    \def\subfigverlabelsize{0.16\textwidth}
    \def\scalefonesize{0.9}
    \def\scalefonesizetitle{0.9}

    \centering
    
    \begin{tikzpicture}
        \node [rotate=90, text width=0.2\textwidth] {{\scalefont{\scalefonesize} \hspace{1.1cm} Success}};
    \end{tikzpicture}  
    \includegraphics[width=0.7\textwidth]{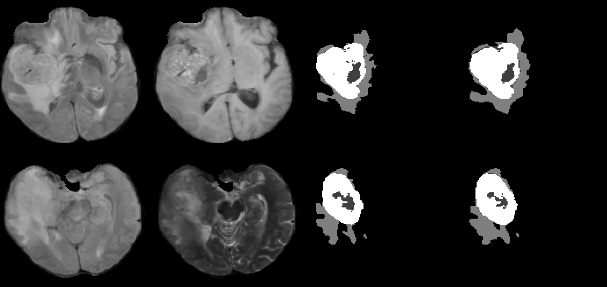} 
    \\

    \vspace{-0.1cm}
    \begin{tikzpicture}
        \node [rotate=90, text width=0.2\textwidth] {{\scalefont{\scalefonesize} \hspace{1.1cm} Failure}};
    \end{tikzpicture}  
    \includegraphics[width=0.7\textwidth]{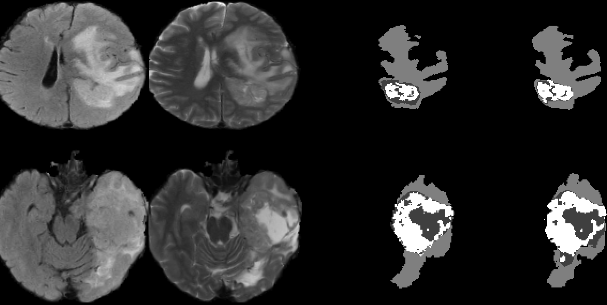} 

    \begin{tikzpicture}
        \node [rotate=90, text width=0.0] {~};
    \end{tikzpicture}
    \captionsetup[subfigure]{labelformat=empty, justification=centering}
    \begin{subfigure}{\subfigsize}
        \caption{{\scalefont{\scalefonesizetitle} FLAIR}}
    \end{subfigure} 
    \begin{subfigure}{\subfigsize}
        \caption{{\scalefont{\scalefonesizetitle} T2}}
    \end{subfigure} 
    \begin{subfigure}{\subfigsize}
        \caption{{\scalefont{\scalefonesizetitle} Ground truth}}
    \end{subfigure}    
    \begin{subfigure}{\subfigsize}
        \caption{{\scalefont{\scalefonesizetitle} Prediction}}
    \end{subfigure} 
    
    \caption[]{A comparison of the ground truth masks and the results. The two examples show the input \gls{t1c} (left), the ground truth masks (middle) and the results of the proposed method (right).}
    \label{fig:qualitative}
\end{figure}

\begin{figure}[!t]
    \begin{center}
        \includegraphics[width=0.7\textwidth]{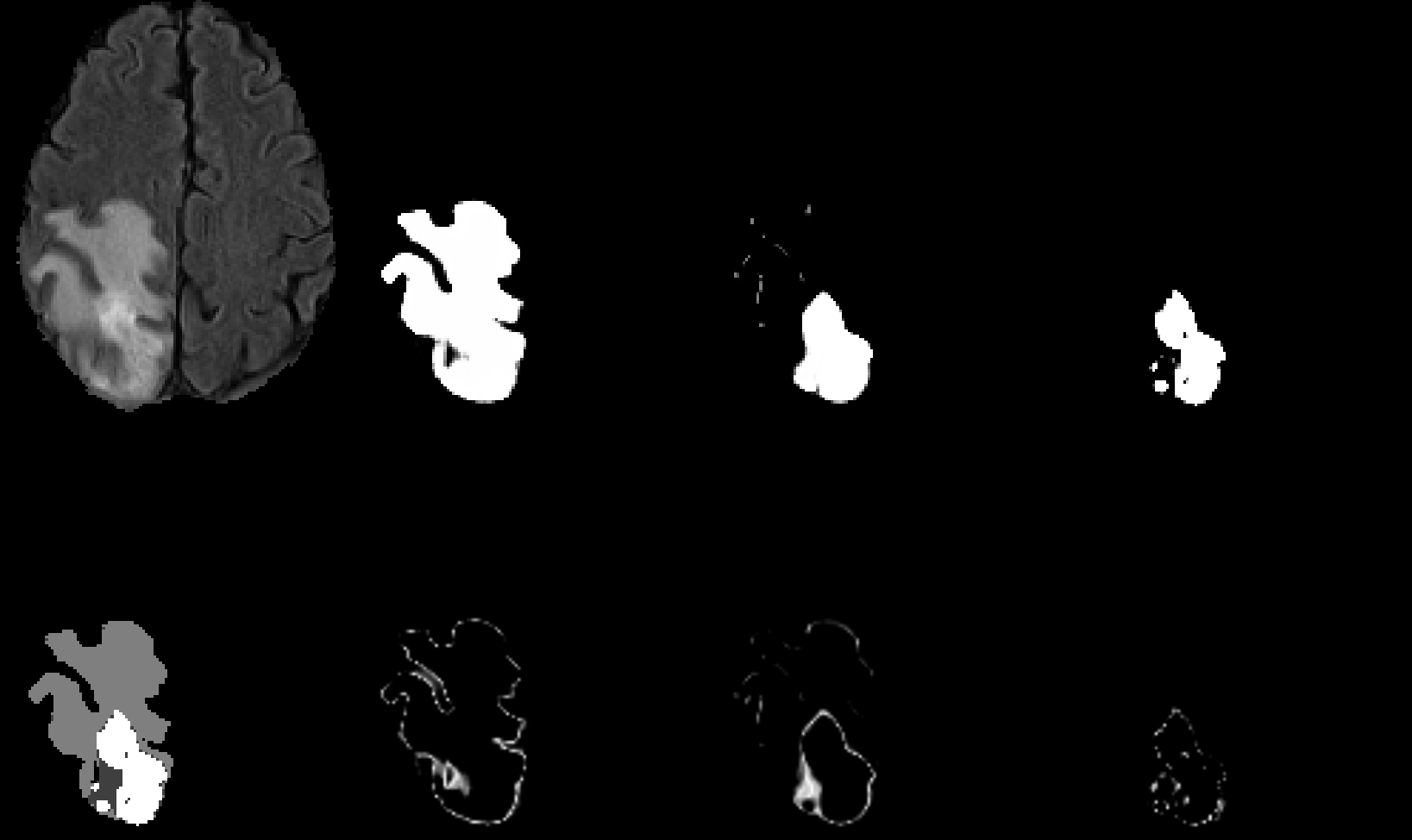}
    \end{center}
    \caption[]{The uncertainty score maps generated from probability maps corresponding to whole, core, and enhancing tumor regions. Top row (from left to right): input, probability map of the whole, core and enhancing tumor regions, respectively. Bottom row (from left to right): multi-class label map (output), uncertainty map of the whole, core and enhancing tumor regions, respectively.}
    \label{fig:unc}
\end{figure}

\tableref{tab:valid} shows the mean \gls{dsc} and \gls{hd95} scores on the validation set by uploading our predicted masks to the evaluation server\footnote{\url{https://www.cbica.upenn.edu/BraTS19/lboardValidation.html}} (team name \textit{UmU}). What is interesting in this table is: (1) \gls{dsc} of core region on the validation set was lower compared to the cross-validation score on the training set, and (2) \gls{tunet} + \gls{seb}, perhaps surprisingly, performed slightly better than the ensemble model of \gls{tunet} + \gls{seb}, \gls{tunet} + \gls{res}$^1$ + \gls{seb} and \gls{tunet} + \gls{res}$^2$ + \gls{seb}. \tableref{tab:test} shows that our BraTS19 testing dataset results are 88.06, 80.84 and 80.29 average dice for whole tumor, tumor core and enhanced tumor core, respectively.





\tableref{tab:valid_unc} provides the mean \gls{dauc} and \gls{rftp} scores on the validation set obtained after uploading our predicted masks and corresponding uncertainty maps to the evaluation server\footnote{\url{https://www.cbica.upenn.edu/BraTS19/lboardValidationUncertainty.html}}. Similar to \tableref{tab:valid}, it can be seen in \tableref{tab:valid_unc} that \gls{tunet} + \gls{seb} is the best-performing model of our models. Though our best-performing model performs slightly worse than the top-ranking teams on the validation set, it was one of the top-performing models on the test set 
This indicates that the proposed ensemble model might generalize the problem well.

\figref{fig:qualitative} illustrates four qualitative examples generated from the Ensemble + post-process model on the training set. As can be seen in the first and second rows, our model detected all three regions well. However, it struggled to correctly differentiate between the necrotic and non-enhancing tumor core and the enhancing tumor (third and last rows). This is most likely due to difficulties in labeling the homogeneous areas.

\figref{fig:unc} illustrates the uncertainty score maps generated from the corresponding output probability maps, that are produced by the \gls{tunet}, for three tumor regions (whole, core, and enhancing). As can be seen in \figref{fig:unc}, the uncertainty scores tends to be: (i) more obvious at the borderline or overlapping areas between tumor regions, and (ii) less apparent on the background or non-tumor regions.

An adapted version of the \gls{tunet} was also used by the authors in the \gls{kits}~\cite{kits19}.

\section{Conclusion}
\label{sec:conclusion}

In conclusion, we developed a cascaded architecture by connecting three U-Net-like networks to segment glioma sub-regions from multimodal brain \gls{mri} images. We separated the complex brain tumor segmentation into three simpler binary tasks to segment the whole tumor, tumor core, and enhancing tumor core, respectively. Our network used an encoder-decoder structure with ResNet-like blocks and Squeeze-and-Excitation blocks after each convolution and concatenation block. Dice scores on the training set and validation set were 91.92/86.45/78.72 and 90.34/81.12/78.42 for the whole tumor, tumor core, and enhancing tumor core, respectively.



\bibliographystyle{splncs04}
\bibliography{bib}

\end{document}